# Albert Einstein's Methodology

Galina Weinstein

This paper discusses Einstein's methodology. The first topic is: Einstein characterized his work as a theory of principle and reasoned that beyond kinematics, the 1905 heuristic relativity principle could offer new connections between non-kinematical concepts. The second topic is: Einstein's creativity and inventiveness and process of thinking; invention or discovery. The third topic is: Einstein considered his best friend Michele Besso as a sounding board and his class-mate from the Polytechnic Marcel Grossman – as his active partner. Yet, Einstein wrote to Arnold Sommerfeld that Grossman will never claim to be considered a co-discoverer of the Einstein-Grossmann theory. He only helped in guiding Einstein through the mathematical literature, but contributed nothing of substance to the results of the theory. Hence, *Einstein neither considered Besso or Grossmann as co-discoverers* of the relativity theory which he himself invented.

## 1 The Principles of Relativity as Heuristic Principles

### 1.1 Einstein's reply to Ehrenfest

Einstein had a friend, Paul Ehrenfest a Jewish physicist from Vienna. In 1907 Ehrenfest wrote a paper.[1] There were the known problems in the 19$^{th}$ century electrodynamics of moving bodies. Einstein's 1905 solution appeared to Ehrenfest very similar to Hendryk Antoon Lorentz's solution to these problems, the celebrated theory of the electron. Ehrenfest thought that Einstein's deformed electron (from his 1905 relativity paper) could have been obtained from the good old theory of Lorentz, if we only used the method of deduction. If this was so, Ehrenfest understood that Einstein's theory was nothing but a reformulation of the electrodynamics of Lorentz. Therefore, Einstein's innovation was the following according to Ehrenfest, "In the formulation in which Mr. Einstein published it, Lorentzian relativity electrodynamics is treated rather generally as a closed system."[2]

Einstein commented on Ehrenfest's paper. His 1907 reply, "Comments on the Note of Mr. Paul Ehrenfest" is important for the demarcation between his theory of relativity and Lorentz's ether-based theory. Lorentz's theory and the descendants of Lorentz's theory are not theories of relativity. Einstein characterized his work what would be later called principle of relativity as a theory of principle and reasoned that beyond kinematics, the 1905 heuristic relativity principle could offer new connections between non-kinematical concepts, "The principle of relativity, or more exactly, the principle of relativity together with the principle of the constancy of the velocity of light, is not to be conceived as a 'closed system', in fact, not as a system at all, but merely as a heuristic principle which, when considered by itself, contains only statements about rigid bodies, clocks, and light signals. The theory of relativity provides something additional only in that it requires relations between otherwise seemingly unrelated regularities".[3]

---

[1] Ehrenfest: 'Die Translation deformierbarer Elektronen und der Flächensatz' ", *Annalen der Physik* 23, 1907, pp. 204-205.

[2] Einstein, Albert, "Bemerkungen zu der Notiz von Hrn. Paul Ehrenfest: 'Die Translation deformierbarer Elektronen und der Flächensatz' ", *Annalen der Physik* 23, 1907, pp. 206-208; p. 206.

Eherenfest's query dealt with the structure of the electron: "Accordingly, it [Lorentz's theory in Einstein's formulation] must also be able to provide purely deductively an answer to the question posed by transferring Max Abraham's problem from the rigid electron to the deformable one […]".[4]

Einstein answered Ehrenfest's query by saying that the theory of the motion of an electron is obtained as follows: "one postulates the Maxwell equations for vacuum for space-time coordinate systems. By applying the space-time transformation [Lorentz transformation] derived by means of the system of relativity, one finds the transformation equations for electric and magnetic forces. Using the latter, and applying the space-time transformation, one arrives at the law for the acceleration of an electron moving at arbitrary speed from the law for the acceleration of a slowly moving electron (which is assumed or obtained from experience)".[5]

Einstein explained to Ehrenfest, "We are not dealing here at all with a 'system' in which the individual laws are implicitly contained and from which they can be found by deduction alone, but only with a principle that (similarly to the second law of the thermodynamics permits the relation of certain laws to others".[6]

In his *Autobiographical Notes* from (1946) 1949, Einstein explained this still further:[7] "Gradually I despaired of the possibility of discovering the true laws by means of constructive efforts based on known facts. The longer and the more desperately I tried, the more I came to the conviction that only the discovery of a universal formal principle could lead us to assured results. The example I saw before me was thermodynamics. The general principle was there given in the theorem […the second law of thermodynamics]. How, then could such a universal principle be found?"

**1.2 Theories of Principle and Constructive Theories**

After 1907 Einstein made a distinction between theories of principle, such as thermodynamics and constructive theories, such as statistical mechanics. He characterized the special theory of relativity as a theory of principle, and considered it to be basically complete when the two underlying principles of the theory (the principle of relativity and that of the constancy of velocity of light) were established. All later work would involve development of constructive theories compatible with these basic principles.

---

[3] Einstein, 1907, pp. 206-207.
[4] ("ein heuristisches Prinzip"). Einstein, 1907, p. 206.
[5] Einstein, 1907, p. 207.
[6] Einstein, 1907, p. 207. It was the first time that Einstein compared the relativity principle to the laws of thermodynamics. *CPAE*, Vol. 2, p. 412, note 8.
[7] Einstein, Albert , "*Autobiographisches*"/"Autobiographical notes" In Schilpp, Paul Arthur (ed.), *Albert Einstein: Philosopher-Scientist*, 1949, La Salle, IL: Open Court, pp. 1–95; pp. 48-49.

In his 1916 popular book, *Relativity, the Special and the General Theory*, in the chapter "The Heuristic Value of The Theory of Relativity", Einstein wrote: "[…] the theory becomes a valuable heuristic aid in the search for general laws of nature".[8]

In his paper, "What is the Theory of Relativity?", written at the request of the *London Times* and published on November 28, 1919, for the first time Einstein formulated his views in a systematic manner:[9]

"We can distinguish various kinds of theories in physics. Most of them are constructive. They attempt to build up a picture of the more complex phenomena out of the materials of a relatively simple formal scheme from which they start out. Thus the kinetic theory of gases seeks to reduce mechanical, thermal, and diffusional processes to the movements of molecules – i.e., to build them up out of the hypothesis of molecular motion. When we say that we have succeeded in understanding a group of natural processes, we invariably mean that a constructive theory has been found which covers the processes in question.

Along with this most important class of theories there exists a second, which I will call 'principle theories'. […]

The advantages of the constructive theory are completeness, adaptability, and clearness; those of the principle theory are logical perfection and security of the foundations.

The theory of relativity belongs to the latter class. In order to grasp its nature, one needs first of all to become acquainted with the principles on which it is based".

**2. Invention or Discovery**

Einstein wrote to Michele Besso in 1948:[10] "I see his [Mach's] weakness in his belief more or less that science consists in the mere 'ordering' of empirical material. i.e., he misjudged the free constructive element in the formation of concepts. He believed that in some sense theories arise by discovery and not invention".

---

[8] Einstein, Albert, *Uber die Spezielle und die Allgemeine Relativitätstheorie*, *Gemeinverständlich*, 1920, Braunschweig: Vieweg Shohn, p. 29
[9] Einstein, Albert, *Ideas and Opinions*, 1954, New Jersey: Crown publishers, p. 228; *The London Times*, November 28, 1919. It should be borne in mind that Einstein wrote this article *after* developing the General Theory of Relativity, and when he spoke about the theory of relativity and the principle theory he probably meant both special and general relativity, because he did not write explicitly the word "special".
[10] Einstein to Besso, January 6, 1948, in Einstein, Albert and Besso, Michele, *Correspondence 1903-1955* translated by Pierre Speziali, 1971, Paris: Hermann, Letter 153 (Speziali).

However, recall that at the same time in (1946) 1949, in his *Autobiographical Notes* Einstein said:[11] "Gradually I despaired of the possibility of discovering the true laws by means of constructive efforts based on known facts. The longer and the more desperately I tried, the more I came to the conviction that only the discovery of a universal formal principle […]". This is the conventional English translation by Schillp.

Einstein wrote his *Autobiographical Notes* in German, and in German the above paragraph is the following: "Nach und nach verzweifelte ich an der Möglichkeit die wahren Gesetze durch auf bekannte Tatsachen sich stützende konstruktive Bemühungen herauszufinden. Je länger und verzweifelter ich mich bemühte, desto mehr kam ich zu der Überzeugung, dass nur die Auffindung eines allgemeinen formalen Prinzipes uns zu gesicherten Ergebnissen führen könnte".[12]

Einstein used two words: "*herauszufinden*" (find out): Gradually I despaired of the possibility of finding out the true laws […]"; and: "*Auffindung*" (discovering): "I came to the conviction that only the discovery of a universal formal principle […]". Einstein thus did not find out true laws using constructive efforts. Instead he discovered a universal principle, in this case, the principle of relativity.

Alberto Martínez asks in his latest book *Kinematics*, "Was the formulation of the special theory of relativity a discovery [*Entdeckung*] or an invention [*Erfindung*]? Nowadays, many writers call it a 'discovery'. But throughout his life, Einstein emphasized the importance of invention, when characterizing his theoretical contribution."[13]

John Stachel argues that according to Einstein, the process of thinking consists of two stages. The first stage "invention", is a solitary activity, primary non-verbal in nature. "Many of the crucial thought experiments Einstein later reports confirm the existence of this stage of the thinking process, utilizing visual and muscular imagery" (e.g., chasing a light ray at the speed of light, the magnet and conductor thought experiment). At a secondary stage, it was necessary for him to transform the results of this primary process into forms communicable to others. This led Einstein to search throughout his early life for people to act as "sounding boards" for his ideas. These people were capable of understanding the things that he explained to them, and of asking intelligent questions that could help Einstein develop his own ideas, but were not capable of any creative effort of their own. Einstein moved back and forth between the two stages in the course of the development of his ideas.[14]

---

[11] Einstein, 1949, p. 49.
[12] Einstein, 1949, p. 48.
[13] Martínez, Alberto, *Kinematics. The Lost Origins of Einstein's Relativity*, 2009, Baltimore: The John Hopkins University Press, p. 285.
[14] Stachel, John, *Einstein's Miraculous Year. Five Papers that Changed the Face of Physics*, 1998/2005, Princeton and Oxford: Princeton University Press, p. xxxv, p. xxxviii.

## 3. Moszkowski's Conversations with Einstein on Inventiveness

In the final chapter of his book, *Conversations with Einstein*, *Gesprächen mit Einstein*, Alexander Moszkowski wrote, "At first it staggered me to hear Einstein say" that "The use of the word 'Discovery' in itself is to be deprecated. For discovery is equivalent to becoming aware of a thing which is already formed; this links up with proof, which no longer bears the character of 'discovery' but, in the last instance, of the means that leads to discovery". According to Moszkowski, Einstein "then stated at first in blunt terms, which he afterwards elaborated by giving detailed illustrations" that "Discovery is really not a creative act!". [15]

According to Moszkowski, Einstein told him, "For it is not true that this fundamental principle occurred to me as the primary thought. If this had been so perhaps it would be justifiable to call it a 'discovery'. But the suddenness with which you assume it to have occurred to me must be denied. Actually, I was lead to it by *steps* arising from the *individual* laws derived from experience". Moszkowski then says that Einstein supplemented this by emphasizing the conception "invention" and ascribed considerable importance to it: "Invention occurs here as a constructive act. This does not, therefore, constitute what is essentially original in the matter, but the creation of a method of thought to arrive at a logically coherent system… the really valuable factor is intuition!"[16]

Towards the end of his book, Moszkowski found a close analogy between technological inventiveness and scientific inventiveness; he used the notion "invention" and connected it with patents and inventors (empirical and technical work on machines and inventions). Moszkowski wrote: "In 1901, after living in Switzerland for five years, he acquired the citizenship of Zürich, and this at last gave him the opportunity of rising above material cares. His University friend, Marcel Grossman lent him a helping hand by recommending him to the Swiss Patent Office, the director of which was his personal friend. Einstein occupied himself here from 1902 to 1905 as a technical expert, that is, as an examiner of applications for patents, and this position gave him the chance of moving about in absolute in the realms of technical science. Whoever has a strong predilection for discovery will perhaps feel estranged to find Einstein so long in the sphere of 'invention', but, as Einstein himself emphasizes very strongly, both regions make great demands on clearly defined and accurate thought. He recognizes a definite relationship between the knowledge that he gained at the Patent Office and the theoretical results that appeared at the same time as products of intensive thought". [17]

---

[15] Moszkowski, Alexander (1921b), *Einstein the Searcher His Works Explained from Dialogues with Einstein*, 1921, translated by Henry L. Brose, London: Methuen & Go. LTD; appeared in 1970 as: *Conversations with Einstein*, London: Sidgwick & Jackson, 1970, pp. 94-95; Moszkowski, Alexander (1921a), *Einstein, Einblicke in seine Gedankenwelt. Gemeinverständliche Betrachtungen über die Relativitätstheorie und ein neues Weltsystem. Entwickelt aus Gesprächen mit Einstein*, 1921, Hamburg: Hoffmann und Campe/ Berlin: F. Fontane & Co, p. 100.

[16] Moszkowski, 1921b, p. 96; Moszkowski, 1921a, p. 101.

## 4. Hadamard's conversations with Einstein on his creativity

Einstein used to describe the process by which fundamental laws are obtained as "free creation of the mind". In an unpublished opening lecture for a course on the theory of relativity that Einstein gave in Argentina in 1925, he said, "Not only are fundamental laws the result of an act of imagination that cannot be controlled, but so are their ingredients, the ideas derived from those laws. Thus, the concept of acceleration was in itself an act of free creation of the mind which, even if supported by the observation of the motion of solid bodies, assumes as a precondition nothing less than the infinitesimal calculus".[18]

Jacques Hadamard, while preparing his 1945 book, *The Mathematical Mind*, asked Einstein some questions about the process by which his ideas developed, and published Einstein's answers,[19]

Hadamard asked: "It would be very helpful for the purpose of psychological investigation to know what internal or mental images, what kind of 'internal world' mathematicians make use of; whether they are motor, auditory, visual, or mixed, depending on the subject which they are studying". Einstein replied:[20]

"(A) The words or the language, as they are written or spoken, do not seem to play any role in my mechanism or thought. The psychical entities which seem to serve as elements in thought are certain signs and more or less clear images which can be 'voluntarily' reproduced and combined.

There is of course, a certain connection between those elements and relevant logically connected concepts. […]

(B) The above mentioned elements are, in my case, of visual and some of muscular type. Conventional words or other signs have to be sought for laboriously only in a secondary stage, when the mentioned associative play is sufficiently established and can be reproduced at will.

(C) According to what has been said, the play with the mentioned elements is aimed to be analogous to certain logical connections one is searching for".

---

[17] Moszkowski, 1921b, p. 229; Moszkowski, 1921a, pp. 226-227.
[18] Einstein, Albert, "Unpublished Opening Lecture for the Course on the Theory of Relativity in Argentina, 1925", translated by Alejandro Gangui and Eduardo L. Ortiz, *Science in Context* 21, 2008, pp. 451-459; p. 453.
[19] Hadamard, Jacques, *The Mathematical Mind*, 1945, New Jersey: Princeton Science Library, p. 140.
[20] Hadamard, 1945, pp. 142-143.

Hadamard asked Einstein another question he did not specify in his book, and Einstein's reply was,[21]

"(D) Visual and motor. In a stage when words intervene at all, they are, in my case, purely auditory, but they intervene only in a secondary stage as already mentioned".

Hadamard then asked Einstein,[22]

"Especially in research thought, do the mental pictures or internal words present themselves in the full consciousness or in the fringe-consciousness […]? Einstein's reply was,[23]

"(E) It seems to me that what you call full consciousness is a limit case which can never be fully accomplished. This seems to me connected with the fact called the narrowness of consciousness (Enge des Bewusstseins).

Remark: Professor Max Wertheimer has tried to investigate the distinction between merely associating or combining of reproducible elements and understanding (organisches Begreifen); I cannot judge how far his psychological analysis catches the essential point".[24]

## 5. A Sounding Board and a Scientific Partner

According to Stachel, Einstein ruminated or was brooding on the idea, but while studying at the Polytechnic, he was talking with his sounding board friends as he was reading the works of great physicists. According to Seelig's report, Einstein's best friend Michele Besso was his sounding board. However, Einstein did not seem to consider his classmate from the Polytechnic, Marcel Grossmann, to be his sounding board.

Martínez objects to characterizing Besso as a "sounding board" for Einstein's ideas. Martínez writes that this description is "first used by Einstein but repudiated by Besso as downplaying his role in their discussions and collaborations".[25]

Later in 1913 Besso came to Zürich and actively participated in solving the Einstein-Grossmann ("Entwurf ") gravitation equations with Einstein. They both tried to find solutions to the problem of the advance of the perihelion of Mercury. The young Einstein may have considered Besso as his sounding board, but was Besso still Einstein's sounding board in 1913?

I will discuss this question and analyze three examples: the magnet and conductor thought experiment (special relativity), solving the problem of the Perihelion of

---

[21] Hadamard, 1945, p. 143.
[22] Hadamard, 1945, p. 141.
[23] Hadamard, 1945, p. 143.
[24] Wertheimer, Max, *Productive Thinking*, 1916/1945, New-York: Harper & Brothers.
[25] Martínez, Alberto, "Review of John Stachel's *Einstein from 'B' to 'Z'*", *Physics in Perspective* 5, 2003, pp. 352-354, p. 354.

Mercury (general relativity), and the second stage of the development of the general theory of relativity. Einstein's Odyssey to general relativity went through three stages. In 1920, Einstein wrote a short list of "my most important scientific ideas". The final three items on the list are: [26]

1907 Basic idea for the theory of relativity [*first stage*: 1917 – 1911: Equivalence Principle, coordinate-dependent theory, and theory of static gravitational field]

1912 Recognition of the non-Euclidean nature of the metric and its physical determination by gravitation [*second stage*: 1912 – 1915: Zurich Notebook and the Einstein-Grossman theory]

1915 Field equations of gravitation. Explanation of the perihelion motion of Mercury [*third stage*: 1915-1916: General Theory of Relativity].

**5.1 The Magnet and Conductor Thought Experiment**

Einstein started his paper with the problematic asymmetries that were inherent in the electrodynamical explanation of the phenomenon of induction by Faraday. Thus the magnet and conductor thought experiment opens Einstein's 1905 relativity paper and not the famous Michelson-Morley second order in *v/c* ether drift experiment. In fact this latter experiment is not even mentioned in the relativity paper.

Einstein explained the thought experiment in the following way:[27]

"It is well known that Maxwell's electrodynamics, as usually understood at present, when applied to moving bodies, leads to asymmetries which do not appear to be inherent in the phenomena. Take, for example, the electrodynamic interaction between a magnet and a conductor. The observable phenomenon here depends only on the relative motion of the conductor and the magnet, whereas the customary view draws a sharp distinction between the two cases, in which either the one or the other of these bodies is in motion".

When the magnet is considered to be moving it is induction: [28]

"For if the magnet is in motion and the conductor is at rest, there arises in the vicinity of the magnet an electric field with a certain definite energy, producing a current at the places where parts of the conductor are located".

And this is not induction; it is force on a moving body: [29]

---

[26] Stachel, John, "The First-two Acts", in Stachel, John, *Einstein from 'B' to 'Z'*, 2002, Washington D.C.: Birkhauser, pp. 261-292; p. 261.
[27] Einstein, Albert, "Zur Elektrodynamik bewegter Körper, *Annalen der Physik* 17, 1, 1905, pp. 891-921; p. 891.
[28] Einstein, 1905, p. 891.
[29] Einstein, 1905, p. 891.

"But if the magnet is at rest and the conductor is moving, no electric field arises in the vicinity of the magnet. In the conductor, however, we find an electromotive force, to which there is no corresponding energy in itself, but which gives rise, assuming equality of relative motion in the two cases discussed, to electric currents of the same magnitude and intensity as those produced by the electric forces in the former case".

If the conductor is at rest *in the ether* and the magnet is moved with a given velocity, a certain electric current is induced in the conductor. If the magnet is at rest, and the conductor moves with the same relative velocity, a current of the same magnitude and direction is in the conductor.

However, *the ether theory* gives a different explanation for the origin of this current in the two cases. In the first case an electric field is supposed to be created *in the ether* by the motion of the magnet relative to it (Faraday's induction law). In the second case, no such electric field is supposed to be present since the magnet *is at rest in the ether*, but the current results from the motion of the conductor through the static magnetic field (Lorentz's force law).

This asymmetry of the explanation is foreign to the phenomenon, because the observable phenomena (the current in the conductor) *depend only on the relative motion* of the conductor and the magnet. According to this experiment observable electromagnetic phenomena should depend only on the relative motions of ponderable matter.

Therefore, "Examples of this sort, together with the unsuccessful attempts to detect motion of the earth relative to the 'light medium' [ether drift experiments] lead to the conjecture that the phenomena of electrodynamics as well as those of mechanics possess no properties corresponding to the idea of absolute rest." [30]

Einstein objected to the idea of absolute rest. Einstein wrote: "The introduction of a 'light ether' ['Lichtäthers'] will prove to be superfluous". The ether is not good for anything, and therefore there is no point in keeping a useless concept.

Einstein concluded, for every coordinate system in which the laws of mechanics hold good, i.e., inertial system, the laws of electrodynamics and optics are also valid; and he raised this conjecture (the 'Principle of Relativity'['Prinzip der Relativität']) to the status of a postulate. He introduced the light postulate, light is always propagated in empty space with a definite velocity *c* which is independent of the state of motion of the emitting body, and said that it was only apparently incompatible with the former principle.[31]

---

[30] Einstein, 1905, p. 891.
[31] Einstein, 1905, pp. 891-892.

Einstein summarized in his relativity paper, "These two postulates suffice for the attainment of a simple and consistent theory of the electrodynamics of moving bodies based on Maxwell's theory for bodies at rest".[32]

Einstein started his electromagnetic part of the relativity paper again with induction. He was then led to solve the conflict with which he opened the kinematical part of his relativity paper. In the electrodynamical part he was going to discuss the problem of asymmetry in electromagnetism.

In order to discuss induction Einstein first obtained the transformation equation for the electric and magnetic forces (fields) using the Lorentz transformations that he had obtained in section §3 of the kinematical part, and the principles of relativity and that of the constancy of the velocity of light. Einstein used *a heuristic method* in order to obtain the transformation equation for the electric and magnetic fields:[33]

1) Einstein assumes the Maxwell-Hertz equations for empty space are valid for the system at rest *K*. He writes the components of the electric force vector with respect to the x, y, z axes in K in the form (X, Y, X) and the components of the magnetic force vector as (L, M, N).

2) **He applies to** the Maxwell-Hertz equations in *K* **the Lorentz transformations developed in §3** for the special and temporal derivatives, i.e.,

$$\frac{\partial}{\partial t} \to \frac{\partial}{\partial \tau}, \frac{\partial}{\partial x} \to \frac{\partial}{\partial \xi}, etc \dots$$

by referring the electromagnetic processes to the system of coordinates moving with velocity v relative to *K*; he obtains the transformed Maxwell-Hertz equations as they appear when written in *k*.

3) Next Einstein **referred to the principle of relativity**: "The principle of relativity requires that if the Maxwell-Hertz equations for empty space are valid in system *K*, they are also valid in system k".[34] Therefore, the electric and magnetic force vectors – (X', Y', Z') and (L', M', N') – in the moving system *k* satisfy a set of Maxwell-Hertz equations of the same form as in K. This implies the transformation equations for (X, Y, Z), (L, M, N).

4) Einstein requires that the two systems of equations found for *k* (in step 2 – by applying the transformations in §3, and in step 3, when guided by the principle of relativity) **take the same form as** the Maxwell-Hertz equations for the system *K* in step 1).

---

[32] Einstein, 1905, p. 892.
[33] Einstein, 1905, pp. 907-909.
[34] Einstein, 1905, p. 908.

Einstein thus arrives at the equations corresponding to Lorentz's transformations of the coordinates, equations of transformation for electric and magnetic forces (fields):[35]

$X' = X, Y' = \beta(Y - vN/c), Z' = \beta(Z + vM/c),$
$L' = L, M' = \beta(M + vZ/c), N' = \beta(N - vY/c).$

Einstein now returned to the experiment by which he opened his paper, the magnet and conductor thought experiment. He was convinced that there should be an electric field acting on the conductor, even if it is moving. To show that, he needed the transformation law for the components of the electromagnetic field, which showed that the (**v** x **B**) term can be interpreted as an electric field in the rest frame of the conductor. Einstein then interpreted the six equations he has just obtained in the following way:

In an ether-based electrodynamics the explanation was based on an electromotive force. If a unit point electric charge is in motion in an electromagnetic field, there acts on it, in addition to the electric force, an "electromotive force", $\frac{\vec{v}}{c} \times B.$

Einstein's new relations for (X', Y', Z') are thus interpreted in the following way: "If a unit point electric charge is in motion in an electromagnetic field, the force acting on it is equal to the electric force, which is present at the location of the unit charge, and is obtained by transformation of the field to a coordinate system at rest relative to the electrical unit charge (New manner of expression)".

If the magnet is now in motion, then in an ether-based electrodynamics the explanation is based on "magnetomotive forces".

In Einstein's theory, "electric and magnetic forces do not exist independently of the state of motion of the coordinate systems". Einstein adds, "It is also clear that the asymmetry mentioned in the introduction when considering the currents produced by the relative motion of a magnet and a conductor, disappears."[36]

This is quite new: Just as the Lorentz transformation mixes special and temporal coordinates, these transformations laws mix the electric and magnetic components. That is why for Einstein there is one electromagnetic field. This is the true unification of the two fields. Their breakup is relative to the inertial frame. For Einstein the two fields mix and not just interact. With Maxwell they interact, but they do not mix. Maxwell did not know about the electromagnetic field in the sense of Einstein.

Einstein ends section §6 by mentioning unipolar induction machines,[37]

---

[35] Einstein, 1905, p. 909.
[36] Einstein, 1905, p. 910.
[37] Einstein, 1905, p. 910.

"Furthermore it is clear that the asymmetry mentioned in the introduction as arising when we consider the currents produced by the relative motion of a magnet and a conductor, now disappears. Moreover, questions as to the "seat" of electrodynamic electromotive forces (unipolar machines) have no point".

In a 1920 manuscript, "Fundamental Ideas and methods of the Theory of Relativity, Presented in Their Development", Einstein spoke about the crucial role of the magnet and conductor experiment in his thought,

"When I established the theory of special relativity, the following idea on Faraday's magneto-electric induction – so far not mentioned [in the paper] – played a guiding role for me.

According to Faraday, during the relative motion of a magnet and an electric circuit, an electric current is induced in the latter. Whether the magnet is moved or the conductor doesn't matter; it only depends on the relative motion. But according to the Maxwell-Lorentz theory, the theoretical interpretation of the phenomenon is very different for the two cases:

[…]

The phenomena of magneto-electric induction forced me ["zwang mich"] to postulate the principle of (special) relativity".

Einstein ends his 1905 relativity paper by saying: "In conclusion, I note that when I worked on the problem discussed here, my friend and colleague M. Besso faithfully stood by me, and I am indebted to him for several valuable suggestions.

Bern, Juni 1905 (received 30. June 1905)".[38]

What could Besso's valuable suggestions have been?

Michele Angelo Besso and Einstein were very close friends. Besso was six years older than Einstein. Besso was from a Jewish family; he was born near Zürich, and eventually was brought up in Italy. Besso left for Zürich and enrolled, in October 1891, in the mechanics section of the Zürich Polytechnic. There he took courses from the same professors who later taught Einstein. After four years of brilliant study he obtained his diploma in mechanical engineering and, soon afterwards, a position in an electrical machinery factory in Winterthur, near Zürich. Michele Besso came frequently to Zürich to attend musical soirées – he played the violin, like Einstein, and there he first met Einstein.[39] Toward the end of 1896 or the beginning of 1897, during Einstein's first semester in the Polytechnic, he had met Besso at the Zürich home of a woman named Selina Caprotti, where people would meet to make music on Saturday afternoons. Besso possessed wide knowledge in physics, mathematics and philosophy, and he discussed with Einstein the philosophical foundations of physics.[40] Later Besso worked with Einstein in the Patent Office in Bern, and after Einstein had left the Office for his first academic position, Besso continued to work there. Besso died a month before Einstein.

---

[38] Einstein, 1905, pp. 921.
[39] Speziali, Pierre, "Einstein writes to his best friend", in French, A. P. (ed), *Einstein A Centenary Volume*, 1979, London: Heinemann for the International Commission on Physics Education, pp. 263-269; pp. 263-264.

Einstein's biographer, Carl Seelig, wrote:[41]

"The first friend to hear of the relativity theory was the engineer, Michele Angelo Besso.[…] Since the two civil servants went the same home and Besso was always eager to discuss the subjects of which he knew a great deal – sociology, medicine, mathematics, physics and philosophy – Einstein initiated him into his discovery. Besso immediately recognized it as a discovery of the utmost importance and of the greatest consequence. The main subject of discussion was the discovery of the light quanta. In endless conversations his cultured friend, in the role of a critical disbeliever, defended Newton's recognized time and space concepts, into which he wove Mach's sensualistic positivism, and his analytical criticism of Newtonian mechanics. Later Besso […] used the following analogy: Einstein the eagle has taken Besso the sparrow under his wing. Then the sparrow fluttered a little higher: 'I could not have found a better sounding-board in the whole of Europe' [Einen besseren Resonanzboden hätte ich in ganz Europa nicht finden können], Einstein remarked when the conversation turned one day to Besso. This way Einstein and Besso became inseparable".

In a letter of August 3, 1952, Besso recounted,[42]

"Another little fairy tale of mine concerning my view that I had participated in [the formulation of] the special theory of relativity. It seemed to me, as an electrical engineer, I must have brought up, in conversations with you, the question, within the context of Maxwell's theory, of what is induced in the inductor of an alternator; depending on whether it is at rest or rotating, there is induced in the inductive part an electromotive [i.e., a magnetic] force or a [purely] electric one, as a peculiarly practical anticipation of the relativistic point of view […]. That this somehow still resonates emotionally within me is demonstrated by the confusing sentence structure. May Spinoza and Freud watch over me".

"Ein anderes Märklein: aus meiner Meinung doch an der SRT beteiligt gewesen zu sein: indem mir, als Elektrotechniker, nahe liegen musste, das, was im Rahmen der Maxwellschen Theorie, je nachdem der Induktor eines Alternators ruht oder rotiert, im induzierten Teil als elektromotorische oder als elektrische Kraft auftritt, als eine eigentümliche praktische Vorwegnahme der Relativitätsauffassung – in das Gespräch

---

hineingebracht zu haben... Dass in mir irgendwie noch ein Affekt mitschwingt, zeigt sich am verworrenen Satzbau. Spinoza und Freud halten mich mir gegenüber wach".

Besso first studied mathematics and physics at Trieste, and then at the University of Rome (1891-1895) there he took courses in mathematics and physics, and thus he could have learned Maxwell's theory there, but there is no evidence for this. On the advice of his uncle David, who taught mathematics at the University of Modana (Italy), he left for Zurich and enrolled in October 1891 in the mechanics section of the Federal Polytechnic school, the ETH. After four years of study he obtained his diploma in mechanical engineering, and soon afterwards, a position in an electrical-machinery factory in Zurich. However, he could *not* have been acquainted with Maxwell's theory from his studies in the Polytechnic.[43] Heinrich Friedrich Weber did not teach Maxwell's theory in the Polytechnic.[44] Dr. Joseph Sauter, Weber's assistant and later Einstein's colleague at the Bern Patent Office, recalled that "this theory [Maxwell's] was not yet on the official program of the Zürich Polytechnic School".[45]

Therefore, it was probably Einstein's self-reading about Maxwell's theory, who explained to Besso about this theory. Only after such explanation could Besso "im Rahmen der Maxwellschen Theorie" (within the context of Maxwell's theory) refer to his technical work and speak with Einstein or remind him about induction (of which Einstein had already read about in Dr. Jos Krist's book). Einstein studied Faraday's induction at secondary school, the Luitpold Gymnasium. He studied from a physics book by Krist, *Anfangsgründe der Naturlehre für die Unterclassen der Realschulen*. Krist in this book explained "Magneto-induction", and then associated it with Faraday's induction.[46]

Einstein had probably also read about induction in August Föppl's book. Anton Reiser writes, "Just then his one great love among the sciences was physics. But the scientific courses offered to him in Zürich soon seemed insufficient and inadequate, so that he habitually cut his classes. His development as a scientist did not suffer thereby. With a veritable mania for reading, day and night, he went through the works of the great physicists – Kirchhoff, Hertz, Helmholtz, Föppl".[47] However, it is important to stress that Einstein never mentioned Föppl among his reading list.

---

[43] Speziali, Pierre, "Einstein writes to his best friend", in French, (ed), 1979, pp. 263-269; pp. 263-264.
[44] See my paper: Weinstein, Galina, "Einstein Chases a Light Beam", ArXiv: 1204.1833v1 [physics.hist-ph], 9 April, 2012.
[45] Sauter, Joseph, "Comment j'ai appris à connaître Einstein", 1960, in Flückiger, Max, *Albert Einstein in Bern*, 1974, Switzerland: Verlag Paul Haupt Bern, p. 154.
[46] Krist, Jos. Dr., *Anfangsgründe der Naturlehre für die Unterclassen der Realschulen*, 1891, Wien: Wilhelm Braumüller, K. U. K Hof. Und Universitäts-Buchhändler, p. 94.
Induction is described in Krist's book and then he wrote: "*Diese Induction heißt M a g n e t o - I n d u c t i o n (F a r a d a y 1831)*".
[47] Reiser, Anton, *Albert Einstein, Ein Biographisches Porträt*, 1930, New-York: Albert & Carles Boni, p. 49.
The journalist Rodulf Kayser, Einstein's son-in-law, the husband of Einstein's stepdaughter Ilse (writing under the pseudonym Anton Reiser), wrote a biography with Einstein's approval and his so-called cooperation. However, Einstein did not consider Reiser's biography as a most reliable book.

According to Besso's letter to Einstein from 1952, Besso and Einstein discussed Faradays induction within the context of Maxwell's theory. Einstein considered Besso as his sounding board even though Besso felt that he had participated in the formulation of the special theory of relativity; but Besso wrote to Einstein that this is "another little fairy story [Märklein] of mine".

**5.2 The Second Stage of the General Theory of Relativity**

In 1955 Einstein dedicated a short *Autobiographical Sketch* to the relations with his close friend Marcel Grossmann. In this *Skizze* Einstein told the story of the collaboration with Grossman which led to the Einstein-Grossmann ("Entwurf") theory.

Einstein first described his friendship with Marcel Grossmann,[48]

"In these student days, I developed a real friendship with a fellow student Marcel Grossmann. I solemnly went with him once a week to Café 'Metropol' on the Limmat embankment and talked to him not only about our studies, but also about anything that might interest young people whose eyes are open. He was not a kind of Vagabond and Eigenbrödler [loner] like me. […] Besides this he had just those gifts in abundance, which I lacked: quick learner and ordered in every sense. He not only visited instead of us in all eligible courses, but he also wrote them so neatly that he had printed his notebooks very well. In preparation for the exams he lent me these notebooks, which meant for me a lifesaver; what would have happened with me without it, I would rather not write and speculate".

Einstein acknowledged his debt to Grossmann for helping him again in 1912. Einstein spent a few months working at the Prague German University, and published two papers discussing the theory of the static gravitational field. He was then appointed to the Zurich Polytechnic, and recognized that the gravitational field should not be described by a variable speed of light as he had attempted to do in Prague, but by the metric tensor field; a mathematical object of ten independent components, that characterizes the geometry of space and time.

Einstein wrote about his switch of attitude towards mathematics in the oft-quoted letter to Sommerfeld on October 29, 1912,[49]

---

Einstein wrote in the preface to Reiser's book: "The author of this book is one who knows me rather intimately in my endeavor, thoughts, beliefs – in bedroom slippers […] I found the facts of this book duly accurate, and its characterization, throughout, as good as might be expected of one who is perforce himself, and who can no more be another than I can".

[48] Einstein, Albert, "Erinnerungen-Souvenirs", *Schweizerische Hochschulzeitung* 28 (Sonderheft), 1955, pp. 145-148, pp. 151-153; Reprinted as, "Autobiographische Skizze" in Seelig Carl, *Helle Zeit – Dunkle Zeit. In memoriam Albert Einstein*, 1956, Zürich: Branschweig: Friedr. Vieweg Sohn/Europa, pp. 9-17; p. 11.

[49] Einstein to Sommerfeld, October 29, 1912, *The Collected Papers of Albert Einstein (CPAE) Vol. 5: The Swiss Years: Correspondence, 1902–1914*, Klein, Martin J., Kox, A.J., and Schulmann, Robert (eds.), Princeton: Princeton University Press, 1993, Doc. 421.

"I am now occupied exclusively with the gravitational problem, and believe that I can overcome all difficulties with the help of a local mathematician friend. But one thing is certain, never before in my life have I troubled myself over anything so much, and that I have gained great respect for mathematics, whose more subtle parts I considered until now, in my ignorance, as pure luxury! Compared with this problem, the original theory of relativity is childish".

After arriving back to Zurich in summer 1912 Einstein was searching his "local mathematician friend" from collage, Marcel Grossmann,[50]

"With this task in mind, in 1912, I was looking for my old student friend Marcel Grossmann, who had meanwhile become a professor of mathematics in the Swiss Federal Polytechnic institute. He was immediately caught in the fire, even though he had as a real mathematician a somewhat skeptical attitude towards physics.

So he arrived and he was indeed happy to collaborate on the problem, but with the restriction that he would not be responsible for any statements and won't assume any interpretations of physical nature".

Einstein's collaboration with Marcel Grossmann led to two joint papers: the first of these was published before the end of June 1913,[51] and the second, almost a year later,[52] two months after Einstein's move to Berlin.[53]

Einstein and Grossmann's first joint paper entitled, "Entwurf einer verallgemeinerten Relativitätstheorie und einer Theorie der Gravitation" ("Outline of a Generalized Theory of Relativity and of a Theory of Gravitation") is called by scholars the "Entwurf" paper. Grossmann wrote the mathematical part of this paper and Einstein wrote the physical part. The paper was first published in 1913 by B. G. Teubner (Leipzig and Berlin). And then it was reprinted with added "Bemerkungen" (remark) in the *Zeitschrift für Mathematik und Physik* in 1914. The "Bemerkungen" was written by Einstein and contained the well-known "Hole Argument".[54] Einstein left Zurich in March-April 1914, and by this ended his collaboration with Marcel Grossmann.

The "Entwurf" theory was already very close to Einstein's general theory of relativity that he published in November 1915. In this theory the gravitational field is represented by a metric tensor, the mathematical apparatus of the theory is based on the work of Riemann, Christoffel, Ricci and Levi-Civita on differential covariants, and the action of gravity on other physical processes is represented by generally covariant equations (that is, in a form which remained unchanged under all coordinate transformations). However, there was a difference between the two theories, the Einstein-Grossmann "Entwurf" and Einstein's 1915 general relativity. The "Entwurf" theory contained different field equations that represented the gravitational field, and these were not generally covariant.[55]

---

[50] Einstein, 1955, pp. 15-16.
[51] Einstein, Albert, and Grossmann, Marcel, *Entwurf einer verallgemeinerten Relativitätstheorie und einer Theorie der Gravitation I. Physikalischer Teil von Albert Einstein. II. Mathematischer Teil von Marcel Grossman*, 1913, Leipzig and Berlin: B. G. Teubner. Reprinted with added "Bemerkungen", *Zeitschrift für Mathematik und Physik* 62, 1914, pp. 225-261.
[52] Einstein, Albert and Grossmann, Marcel, "Kovarianzeigenschaften der Feldgleichungen der auf die verallgemeinerte Relativitätstheorie gegründeten Gravitationstheorie", *Zeitschrift für Mathematik und Physik* 63, 1914, pp. 215-225.
[53] *The Collected Papers of Albert Einstein, Vol. 4: The Swiss Years: Writings, 1912–1914*, Klein, Martin J., Kox, A.J., Renn, Jürgen, and Schulmann, Robert (eds.), Princeton: Princeton University Press, 1995, "Einstein on Gravitation and Relativity: The Collaboration with Marcel Grossman", p. 294.
[54] Einstein and Grossman, 1913.

Let us go back to 1912. How did Einstein recall his interaction with Grossmann? Einstein came back to Zurich, Grossmann helped him in his search for a gravitational tensor. Einstein stressed in the *Skizze* that Grossmann,[56]

"He looked through the literature, and soon discovered that the particular implied mathematical problem was already solved by Riemann, Ricci and Levi-Civita. The entire development followed the Gaussian theory of curvature-surfaces, which was the first systematical use of generalized coordinates. Reimann's achievement was the biggest. He showed how a field of *gik* tensors of the second differentiation rank can be formed".

Grossmann brought to Einstein's attention the works of Riemann, Ricci and Levi-Civita; and in addition he showed him the work of Christoffel. Louis Kollros, a professor of geometry and mathematics at the ETH, who was a fellow student of Einstein, wrote in memory of Einstein in 1955 that, sometime upon his arrival, Einstein spoke about his concern with Grossmann and told him one day, "Grossmann, you have to help me, or I shall go crazy! And Marcel Grossman had managed to show him that the mathematical tools he needed, were created just in Zurich in 1869 by Christoffel in the treatise 'On the Transformation of the Homogeneous differential Forms of the Second Degree', published in Volume 70. of the 'Journal de Crelle' for pure and applied mathematics".[57]

Just before writing the 1913 "Entwurf" paper with Grossmann, Einstein had struggled with these new tools in a small blue Notebook – named by scholars the "Zurich Notebook".[58] Einstein filled 43 pages of this notebook with calculations, while he was fascinated with Riemann's calculus. While filling the notebook he received from time to time the new mathematical tools from Grossmann, and he wrote Grossmann's name in the notebook every time he got something new to indicate the tensors that he received from him. At the top of the 14L page Einstein wrote on the left: "Grossmann's tensor four-manifold" and next to it on the right he wrote the fully covariant form of the Riemann tensor. On top of the 22R page he wrote Grossmann's name.[59] Einstein considered on page 22R candidate field equations with a gravitational tensor that is constructed from the Ricci tensor; an equation Einstein would come back to in his November, 4 1915 paper on General Relativity.

---

[55] *CPAE*, Vol. 4, "Einstein on Gravitation and Relativity: The Collaboration with Marcel Grossmann", p. 294.
[56] Einstein, 1955, p. 16.
[57] Kollros, Louis, "Erinnerungen eines Kommilitonen", 1955, in Seelig Carl, *Helle Zeit – Dunkle Zeit. In memoriam Albert Einstein*, 1956, Zürich: Branschweig: Friedr. Vieweg Sohn/Europa, p. 27.
[58] *CPAE*, vol. 4, Doc. 10.

Einstein wrote in the introduction to this November 4 paper, "I completely lost trust in my established field equations [of the Einstein-Grossmann "Entwurf" theory], and looked for a way to limit the possibilities in a natural manner. Thus I arrived back at the demand of a broader general covariance for the field equations, from which I parted, though with a heavy heart, three years ago when I worked together with my friend Grossmann. As a matter of fact, we then have already come quite close to the solution of the problem given in the following".[60]

Einstein wrote towards the end of the *Skizze*,[61]

"While I was busy at work together with my old friend, none of us thought of that tricky suffering, now this noble man is deceased. The courage to write this little colorful Autobiographical Sketch, gave me the desire to express at least once in life my gratitude to Marcel Grossmann".

On July 15, 1915, Einstein gave the clearest statement about Grossmann's role in the development of the (second stage of the) general theory of relativity, in a letter to Arnold Sommerfeld: "Grossman will never claim to be considered a co-discoverer. He only helped in guiding me through the mathematical literature, but contributed nothing of substance to the results."[62] Einstein thus explained that the Einstein-Grossmann "Entwurf" theory was his own theory; and three months later Einstein took full responsibility, and eventually he blamed only himself, when the "Entwurf" field equations collapsed.

**5.3 The Perihelion of Mercury**

During a visit by Besso to Einstein in Zurich in June 1913 they both tried to solve the "Entwurf" field equations to find the perihelion advance of Mercury in the field of a static sun in what is known by the name, the "Einstein-Besso manuscript".[63] Besso was inducted by Einstein into the necessary calculations. The "Entwurf" theory predicted a perihelion advance of about 18" per century instead of 43" per century.[64]

---

[59] *CPAE*, Vol.4, "Einstein on Gravitation and Relativity: The Collaboration with Marcel Grossmann", p. 296.
[60] Einstein, Albert, "Zur allgemeinen Relativitätstheorie", *Königlich Preußische Akademie der Wissenschaften* (Berlin). *Sitzungsberichte*, 1915, pp. 778-786; p. 778.
[61] Einstein, 1955, p. 16.
[62] "Grossmann wird niemals darauf Anspruch machen, als Mitentdecker zu gelten. Er half mir nur bei der Orientierung über die mathematische Litteratur, trug aber materiell nicht zu den Ergebnissen [be]i". Einstein to Sommerfeld, July 15, 1915, *The Collected Papers of Albert Einstein, Vol. 8: The Berlin Years: Correspondence, 1914–1918* (*CPAE* 8), Schulmann, Robert, Kox, A.J., Janssen, Michel, Illy, Jószef (eds.), Princeton: Princeton University Press, 2002, Doc. 96.
[63] *CPAE*, Vol. 4, Doc. 14.
[64] *CPAE*, Vol. 4, "The Einstein-Besso Manuscript on the Motion of the Perihelion of Mercury", p. 351.

As chance would have it, in the same month Einstein had another visitor. In June 1913 Paul and Tatyana Ehrenfest came from Leyden to a trip to Zurich. They stayed in a pension in Zurich, but they spent a great deal of time with Einstein and his family. What a coincidence that Ehrenfest met Marcel Grossmann during this stay in Zurich, and he also met Besso.[65] Einstein was thus surrounded by Besso and Ehrenfest while he tried to find solutions to the Einstein-Grossmann field equations and solve the problem of the precession of the perihelion of Mercury.

Besso collaborated with Einstein on the wrong gravitational (the "Enrwurf") theory, and their calculation based on this theory gave a wrong result. Towards the end of 1915 Einstein abandoned the Einstein-Grossmann "Entwurf" theory; he transferred the basic framework of the calculation from the Einstein-Besso manuscript, and corrected it according to his new 1915 generally covariant field equations. With his new 1915 General Relativity Theory he got the correct precession so quickly, because he was able to apply the methods he had already worked out two years earlier with Besso. [66] Einstein though did not acknowledge his earlier work with Besso, and did not mention his name in his 1915 paper that explains the anomalous precession of Mercury.

It appears that Einstein did not mention Besso because he still considered him as a sounding board even though Besso was calculating with Einstein in Zurich; this as opposed to his other friend, Marcel Grossmann, who was his active partner since 1912 in creating the "Entwurf" theory. Yet, recall that Einstein wrote to Sommerfeld that Grossman will never claim to be considered a co-discoverer of the Einstein-Grossmann theory. He only helped in guiding Einstein through the mathematical literature, but contributed nothing of substance to the results of the theory.

Indeed when Einstein wrote Besso a series of letters between 1913 and 1916, and described to him step by step his discoveries of General Relativity, Besso indeed functioned again as the good old sounding board as before 1905.[67] After completing the 1915 general theory of relativity, Einstein summarized the situation to Besso,[68] "Ich sandte Dir heute die Arbeiten. Die kühnsten Träume sind nun in Erfüllung gegangen. Allgemeine Kovarianz. Perihelbewegung des Merkur wunderbar genau. [...]". Einstein sent Besso his 1915 general relativity papers, and told his best sounding board friend that his wildest dreams have now come true: general covariance and the perihelion of Mercury.

---

[65] Klein, Martin, J., *Paul Ehrenfest, Volume 1 The making of a Theoretical Physicist*, 1970/1985, Amsterdam: North-Holland, p. 294.
[66] *CPAE*, Vol. 4, Doc. 14, pp. 1-14; *CPAE*, Vol. 4, "The Einstein-Besso Manuscript on the Motion of the Perihelion of Mercury", pp. 349-351.
[67] Speziali, Letters 9 to 14.
[68] Einstein to Besso, Decmeber 10, 1915, *CPAE*, Vol. 8, Doc. 162.

Martínez wrote that a sister of Besso once asked Einstein why Besso had not made comparable achievements:[69]

"Herr Professor", she asked… "this I really meant to ask you for a long time – why hasn't Michele made some important discovery in mathematics?"

"But Frau Bice," said Einstein, laughing "this is a very good sign. Michele is a humanist, a universal spirit, too interested in too many things to become a monomaniac. Only a monomaniac gets what we commonly refer to as results".

In conclusion, Einstein thanked Michele Besso at the end of his 1905 relativity paper "for several valuable suggestions". When Einstein walked back home from the Patent Office with Besso, the latter functioned as Einstein's sounding board.

In 1912 Marcel Grossmann brought to Einstein's attention the works of Riemann, Ricci and Levi-Civita. Einstein struggled with these new tools in the "Zurich Notebook". He considered an equation he would come back to in the November 4, 1915 General Theory of Relativity. In 1913 Einstein's collaboration with Grossmann led to the Einstein-Grossmann theory.

During a visit by Besso to Einstein in 1913 they both tried to solve the Einstein-Grossmann field equations to find the perihelion advance of Mercury. The theory predicted a wrong perihelion advance.

Towards the end of 1915 Einstein abandoned the Einstein-Grossmann theory, and with his new General Theory of Relativity got the correct precession so quickly because he was able to apply the methods he had already worked out two years earlier with Besso. Einstein did not mention Besso, probably because he still considered him as a sounding board; this as opposed to his other friend, Grossmann, who was his active partner since 1912 in creating the Einstein-Grossmann theory.

Yet, Einstein wrote to Sommerfeld that Grossman will never claim to be considered a co-discoverer of the Einstein-Grossmann theory. He only helped in guiding Einstein through the mathematical literature, but contributed nothing of substance to the results of the theory. Hence, *Einstein neither considered Besso or Grossmann as co-discoverers* of the relativity theory which he himself invented.

*I wish to thank Prof. John Stachel from the Center for Einstein Studies in Boston University for sitting with me for many hours discussing special relativity and its history.*

---

[69] Martínez, 2009, p. 290.